\documentclass[twoside]{ilcws08}
\usepackage[latin1]{inputenc}
\usepackage[dvips]{graphicx,epsfig,color}
\usepackage{wrapfig,rotating}
\usepackage{amssymb,amsmath,array}

\begin{document}
\title{
M.i.p. detection performances of a 100 $\mu s$ read-out
CMOS pixel sensor with digitised outputs} 
\author{Marc Winter$^1$, J\'er\^ome Baudot$^1$, Auguste Besson$^1$, 
Claude Colledani$^1$, Yavuz Degerli$^2$, Rita De Masi$^1$, \\
Andr\'ei Dorokhov$^1$, Guy Dozi\`ere$^1$, Wojciech Dulinski$^1$, Marie G\'elin$^2$, 
Fabrice Guilloux$^2$, Abdelkader \\
Himmi$^1$, 
Christine Hu-Guo$^1$, Fr\'ed\'eric Morel$^1$, Fabienne Orsini$^2$,
Isabelle Valin$^1$ 
and Georgios Voutsinas$^1$
\thanks{Work supported by the European project EUDET(FP6).}
\vspace{.3cm}\\
1- Institut Pluridisciplinaire Hubert Curien (IPHC) \\
23 rue du loess - BP 28 - 67037 Strasbourg - France
\vspace{.1cm}\\
2- CEA SACLAY - DSM/IRFU/SEDI \\                  
91191 Gif-sur-Yvette Cedex, France \\
}

\maketitle

\begin{abstract}
  Swift, high resolution CMOS pixel sensors are being developed 
for the ILC vertex detector, aiming to allow approaching the 
interaction point very closely. A major issue is the time resolution 
of the sensors needed to deal with the high occupancy generated 
by the beam related background. 
A 128x576 pixel sensor providing digitised outputs at a read-out 
time of 92.5 $\mu s$, was fabricated in 2008 within the EU project
EUDET, and tested with charged particles at the CERN-SPS. Its 
prominent performances in terms of noise, detection efficiency 
versus fake hit rate, spatial resolution and radiation tolerance 
are overviewed. They validate the sensor architecture.

\end{abstract}

\section{Introduction}

   The ambitionned impact parameter resolution at the ILC
calls for a highly granular and transparent pixellised vertex 
detector with an inner radius well below 2 centimetres. 
Approaching the interaction point so closely exposes the pixel
sensors equipping the detector to high rates of beam-related 
background, dominated by beamstrahlung electrons. The 
latter govern the time resolution and the radiation tolerance 
required for the sensors. 

   CMOS pixel sensors are attractive for this application 
because of their high granularity and thin sensitive volume. 
Adapting them to the ILC vertex detector drives an R\&D 
effort mainly oriented towards a high read-out speed, while 
ensuring that the sensors tolerate the radiation level 
foreseen. 

   IPHC-Strasbourg and IRFU-Saclay are developing such sensors 
within the European FP6 project EUDET~\cite{eudet}, to equip a 
pixellised beam telescope providing a few micrometer resolution 
over each of its six, 1x2 cm$^2$ large, planes read out at a 
frequency of 10$^4$ frames per second. Details on the sensor
design are provided in \cite{pixel08}.

   To accommodate the data rate, the signals delivered by 
the sensors are discriminated before being filtered by an 
integrated zero-suppression logic. A fast read-out is achieved 
by grouping the pixels composing the sensitive area in columns 
read out in parallel. The development of this architecture relies 
on two parallel tasks. One of them addresses the upstream part 
of the signal collection and conditionning chain, including the 
pixel array and the discriminators ending the columns. The other 
concerns the downstream part, combining a zero-suppression 
logic with output memories. 

\section{Pixel array with binary outputs}

   Small prototypes were fabricated and tested in previous years to
develop the upstream part of the sensor architecture~\cite{mimosa16}.
{\sl MIMOSA-22} is the last prototype of this R\&D line before 
realising the final sensor for the EUDET telescope. Two 
complementary versions were designed and fabricated~\cite{pixel08}. 
They feature 136 columns 
read out in parallel, each made of 576, 18.4 $\mu m$ pitch, 
pixels. 128 columns are ended with a discriminator, while 8 
columns have analogue outputs for test purposes. The chip 
incorporates a JTAG controller. The frame read-out time is 
92.5~$\mu s$. Various pixel designs were integrated in the chip, 
allowing to explore different sensing diode sizes, amplification 
schemes, ionising radiation tolerant designs, etc. 

  The sensors were characterised in 2008, first in the laboratory 
with an 
$^{55}$Fe source and next at the CERN-SPS, mounted on a 
silicon-strip beam telescope. A modest noise value was found 
for most pixel designs, ranging from about 10 to 14 e$^-$ENC
at room temperature, with a mild operating temperature 
dependence. 
The 128 discriminators exhibited a modest 
threshold dispersion ($\pm$ 4 \% standard deviation) and 
contributed marginally to the total noise. No significant 
non-uniformity was found over the sensitive area of any of 
the 6 sensors tested.

\begin{wrapfigure}{r}{0.585\columnwidth}
\centerline{\includegraphics[width=0.56\columnwidth]
{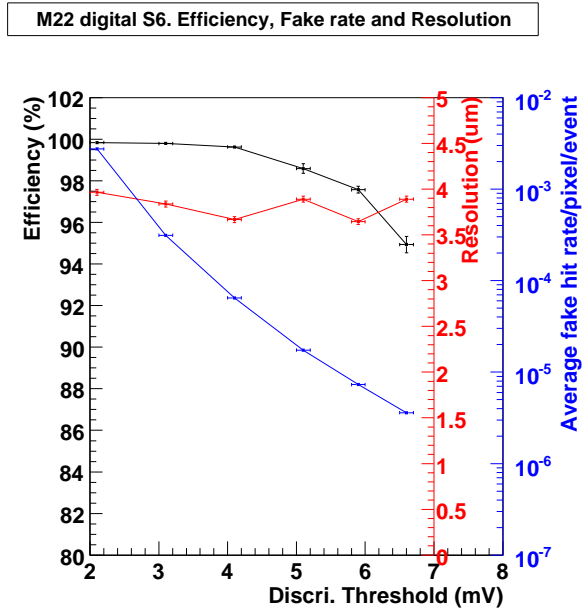}}
\vspace*{-0.4cm}
\caption{Mimosa-22 beam test results.}
\label{Fig:M22}
\end{wrapfigure}
  When exposed to a 120 GeV $\pi^-$ beam at the CERN-SPS,
a signal-to-noise ratio in the  range 17--21 (most probable value)
was measured, depending on the pixel design. Figure \ref{Fig:M22}
displays the measured detection performances of one sub-array
(detection efficiency, single point resolution, average fake hit rate )
as a function of the discriminator thresholds.

   The observed detection efficiency remains \mbox{$>$ 99.5 \%} 
for threshold values high enough to keep the fake hit rate 
\mbox{$<$ 10$^{-4}$}, avoiding the signal processing 
micro-circuits to be saturated by pixel noise fluctuations. 
The single point resolution is $<$ 4 $\mu m$, well suited to the 
detector outer layers and close to the value required for the 
inner layers ($\lesssim$ 3 $\mu m$). Reducing the pitch by 
$\sim$~20\% will suffice to reach this goal.   

   These appealing results were complemented with a 
first study of the sensor tolerance to ionising radiation. The 
expected annual dose at the ILC is $\sim$ O(1) kGy in the 
innermost layer, due to beamstrahlung electrons. 
The detection performances of a sensor exposed to 1.5 kGy 
(10 keV X-Ray source )
were assessed at the CERN-SPS with the 120 GeV $\pi^-$ 
beam mentionned earlier. Despite a $\sim$ 30 \% noise increase,
the detection efficiency was still above 99.5 \% for a fake rate
\mbox{$\lesssim$ 10$^{-4}$}. The noise increase is due to the amplification 
circuitry integrated in the pixels. Its design is being improved, 
aiming for $\sim$ 10 kGy tolerance.

   Summarising, the MIMOSA-22 architecture is validated for 
its integration in the final EUDET sensor.


\section{Zero-suppression micro-circuit}

  {\sl SUZE-01} incorporates the zero-suppression micro-circuit 
and output memories composing the downstream part of the 
sensor architecture. Its logic is intended to be integrated right 
after the discriminators of MIMOSA-22 in the final sensor. 

  The circuitry is organised in a 3 stage pipeline, adapted to 
the output of 128 columns~\cite{pixel08}. In the first stage, the 
input signals 
(discriminator outputs) are distributed over banks, each 
encompassing 64 columns, where a sparse data scan 
algorithm on hit pixels is performed. Up to 4 contiguous pixel 
signals (called "string" hereafter) above threshold are encoded 
in a 2 bit state word. Up to 6 strings per bank can be memorised 
with column addresses. The address of a string shared by two 
neighbouring banks is transferred only once. The second stage
combines the outcomes of the two banks of the first stage. 
Its multiplexing logic accepts up to 9 strings per pixel row and 
adds the bank address information. 
The results of the second stage are stored in the third stage, 
i.e. a 96 kbit memory split in 2 buffers, allowing a continuous 
read-out via a LVDS link at 160 MHz. 

    Fabricated in 2007, SUZE-01 was tested extensively at IPHC 
with millions of patterns at its nominal clock frequency (100 MHz) 
and above. No failures were spotted for frequencies $\leq$ 
115 MHz. The architecture is thus validated for its integration in 
the final EUDET sensor.

    The EUDET telescope specifications, are quite similar to 
those of the ILC vertex detector outer layers. For the inner 
layers, the maximal number of strings per stage, the memory 
capacity and the transfer frequency still need to be extended.

\section{Full size pixel array with integrated zero suppression}

   The final EUDET sensor (called {\sl MIMOSA-26}) was 
designed in 2008 and sent for fabrication. 
It combines the architecture of {\sl MIMOSA-22} and {\sl SUZE-01} 
in a comprehensive charge sensing and signal read-out chain, 
providing discriminated signals in a binary mode including the 
pixel address. It features 1152 columns of 576 pixels, read out 
in $\sim$ 100 $\mu s$. It will be commissionned in 2009, mounted 
on the EUDET telescope. If satisfactory, this architecture will 
next be adapted to the outer layers of the 
vertex detector, which require about 100 $\mu s$ read-out time.  
It will also be evolved towards shorter read-out times, aiming 
for a target value of $\sim$ 25 $\mu s$ for the innermost layer.


\begin{footnotesize}

\end{footnotesize}


\end{document}